\documentstyle[12pt]{article}
\begin{document}
\setlength{\baselineskip}{0.30in}

\newcommand{\beq}{\begin{equation}}
\newcommand{\eeq}{\end{equation}}

\newcommand{\bi}{\bibitem}
\def\mpl{m_{Pl}}

\begin{flushright}
UM-TH-94-19\\
DPNU-94-19\\
June, 1994\\
hep-ph/9406373
\end{flushright}

\begin{center}
\vglue .06in
{\Large \bf {Extended Nambu-Jona-Lasinio Model\\
 vs\\
  QCD Sum Rules}}\\[.5in]

{\bf Koichi Yamawaki}
\\[.05in]
{\it Department of Physics\\
Nagoya University\\
Nagoya, 464-01, Japan}\\
{\bf and}\\
{\bf V.I. Zakharov}\footnote{On leave of absence from Max-Planck
Institute for Physics, Munich}\\
{\it{The Randall Laboratory of Physics\\
University of Michigan\\
Ann Arbor, MI 48109}}\\[.15in]
\end{center}
%{Abstract}\\[-.1in]
\begin{abstract}
\begin{quotation}
We argue that the extended Nambu-Jona-Lasinio model of hadrons can be
be probed and constrained in a nontrivial way
via QCD sum rules. While there arise rather restrictive bounds
on the strength of the effective four-fermion interaction in the
vector channel, introduction of the four-fermion interaction in the
pseudoscalar channel could resolve a long standing puzzle of QCD sum
rules. We speculate also on possible connection between effective
four-quark interactions describing the low-energy phenomenology and
ultraviolet renormalons of the fundamental QCD.

\end{quotation}
\end{abstract}
\newpage

\section{Introduction}

More than thirty years ago Nambu and Jona-Lasinio proposed a model of
superconductivity type as an effective theory of hadrons at low
energies \cite{nambu}.
Advantages of the model are its simplicity and elucidation of
the mechanism of spontaneous breaking of
the chiral symmetry. Since then the model
has developed into a rich phenomenology of hadrons up to mass scale
of order 1 GeV (for a review see, e.g., Ref.\cite{volkov}). Moreover,
 the model has inspired introduction of similar ideas to describe
 hypothetical interactions at
various mass scales (for a review see, e.g., Ref.\cite{koichi}).

The basic feature of (the extended version) of the NJL model is the
 introduction of effective four-fermion
interactions. The form of the interaction is constrained by
the
chiral invariance
of the underlying fundamental Lagrangian and is parametrized in terms
 of two couplings $G_{S,V}$:
\beq
L_{\rm int}~=~{{G_S}\over 2}\left((\bar
q\lambda^{\alpha}q)^2+(\bar qi\gamma_5\lambda^{\alpha}q)^2\right)-
{{G_V}\over 2}\left((\bar q\gamma_{\mu}\lambda^{\alpha}q)^2+
(\bar q\gamma_{\mu}\gamma_5\lambda^{\alpha}q)^2\right),
\label{one}
\eeq
where $\lambda^{\alpha}$ are the Gell-Mann SU(3)
matrices in the flavour space and the colour indices are suppressed.

Loop integrations with Lagrangian (\ref{one}) are allowed only up to
 an ultraviolet cutoff $\Lambda_{UV}$, so that (\ref{one}) represents
 a low energy effective interaction. To get feeling of the parameters
 involved one may keep in
mind the following estimates (see, e.g., Ref.\cite{volkov}):
\beq
\Lambda_{UV}~\approx~1.25~{\rm GeV},~ G_S~\approx~5~ {\rm GeV}^{-2},
{}~G_V~\approx~10~{\rm GeV}^{-2}
\label{numbers}.
\eeq
The fits may somewhat vary, however. It is worth noting that the
scalar type
interaction is responsible for the spontaneous breaking of the chiral
symmetry,
while the function of the vector type interaction is to generate
spin-1 mesons.
The modern way \cite{bijens}
to confront the NJL model (\ref{one}) with the data is to calculate
the parameters of the low-energy chiral Lagrangians classified in
Ref.\cite{leutwyler} at low energies.

A crucial problem is whether the effective
Lagrangian (\ref{one}) could be derived from
the fundamental QCD Lagrangian. This problem has been addressed in a
 number of papers, see in particular review in Ref.\cite{ball}.
 One may not hope at the
moment to derive analytically the effective Lagrangian, since the QCD
coupling is strong at low energies. Thus the common lore does not
extend too far beyond the one-gluon exchange.

Weakness of the standard one-gluon-exchange picture \cite{ball}
is, to our mind, that it does not give any hint as to why the gluon
 line is harder than the quark ones
 so that introduction of a point like four-quark
interaction could be a reasonable approximation.
To our knowledge the only
mechanism inherent to the perturbative QCD which makes gluon lines
hard is the so-called ultraviolet renormalon \cite{thooft}
(for a review see also \cite{book}).
(For a possibility of the nonperturbative QCD to ``induce''
the $G_S$-type four-fermion interaction see Ref.\cite{kondo}.)
 In terms of the sum rules the ultraviolet renormalon results in
$Q^{-2}$ corrections to the parton-model predictions and the
possibility of
such corrections has been brought to attention only recently
\cite{VZrecent,Brown}. Moreover it was demonstrated that indeed the
 ultraviolet renormalon is related to four-quark
 operators \cite{vainshtein}.
However, there are no theoretical means to relate the renormalon
contribution in different channels.

In this paper we will address ourselves to the problem of confronting
the NJL model with the
fundamental QCD. Our basic observation is that in $\rho$ meson
channel the claimed regions of applicability of the effective theory
and of perturbative QCD in fact overlap. Namely, the QCD sum
rules \cite{SVZ} rely, on one hand, on the perturbative
QCD and, on the other hand, are still
valid at Euclidean mass scale as low as $m_{\rho}^2$. Since,
according to the estimates (\ref{numbers}), the ultraviolet cutoff
 $\Lambda^2_{UV}$ in the NJL
model is substantially larger than $m_{\rho}^2$, the sum rules are
sensitive
to the introduction of the phenomenological piece (\ref{one}).

It might worth emphasizing that generally speaking, the use of the
effective Lagrangian and perturbative approach to fundamental QCD
are justified
in different kinematical regions. Namely, the language of (nearly)
massless quarks and of gluons becomes rigorous at short distances
where the running coupling is small, while the effective Lagrangian
(\ref{one}) describes the large-distance, or the
low-momentum dynamics. It is a very specific interplay
of numbers that allows for a window in kinematical variables where
the effective field theory approach to QCD confronts the perturbative
approach to QCD. We use this to obtain independent information on
$G_{S,V}$.

Our conclusion is that the value of $G_V$ is too big to
be consistent with the sum rules. On the other hand,
the effect of $G_S$ is just about what is needed to resolve a long
standing puzzle of QCD sum rules, that is, the failure of the
standard approach \cite{SVZ} in the pseudoscalar channel
\cite{novikov}. Pursuing this line of
reasoning, we conclude that the phenomenological number
(\ref{numbers}) for $G_V$ is to be an overestimate.
Careful analysis of the data from this point of view could serve as a
crucial test of the proposed phenomenology. Without trying to surpass
the results of such an analysis let us note that, to the best of our
understanding, $G_V\neq 0$ is not needed for the NJL model to be
successful in describing
the low-energy pion physics (see \cite{bijens}).

\section{Sum rules in $\rho$ channel}

We start our discussion with estimating the effect of the
interaction (\ref{one}) on the sum rules in $
\rho$ channel. The sum rules \cite{SVZ} are formulated in terms of
$\Pi^{\rho} (Q^2)$,
\beq
\Pi^{\rho}(Q^2)(q_{\mu}q_{\nu}-g_{\mu\nu}q^2)~=~i\int d^4x e^{iqx}
\langle 0|T\{j_{\mu}^{\rho}(x),j_{\nu}^{\rho}(0)\}| 0\rangle,
{}~~Q^2\equiv-q^2,
\label{corr}
\eeq
where $j_{\mu}^{\rho}$ is the quark current with quantum numbers of
$\rho$:
$$
j_{\mu}^{\rho}~=~{1\over 2}(\bar u\gamma_{\mu}u-\bar d\gamma_{\mu}d).
$$
The function $\Pi^{\rho}(Q^2)$ satisfies once subtracted dispersion
relations
\beq
\Pi^{\rho} (Q^2)~=~{{Q^2}\over{\pi}}\int{
{{R^{I=1}(s)ds}\over {s(s+Q^2)}} },
\eeq
where $R^{I=1}(s)$ is the ratio of the cross section of $e^+e^-$
annihilation
into hadrons with total isospin $I=1$ to that of annihilation into
$\mu^+\mu^-$ pair; in particular, $R^{I=1}(s)$ is contributed
by the $\rho$ meson.

The basic idea of the QCD sum rules \cite{SVZ} is to calculate
$\Pi^{\rho}(Q^2)$ at
large  $Q^2$ using the perturbative QCD and then extrapolate the
result to as
low $Q^2$ as possible. Moreover, it turns out that the advance
towards lowest
$Q^2$ is checked by power corrections in $Q^{-2}$.
The coefficients in front of the powers of $Q^{-2}$ are related to
the quark and gluon condensates. Numerically the power like
corrections set
in around $Q^2\sim 0.5~ {\rm GeV}^2\sim m^2_{\rho}$. More precisely,
the corrections
are relatively small at such $Q^2$ but blow up fast at lower $Q^2$.

While the reader is referred to the original papers and reviews
\cite{SVZ} for the details of the sum rules, here we only mention
that the sum rules are
most successful once applied to the Borel transform $\Pi^{\rho} (M^2)$
of $\Pi^{\rho}(Q^2)$. The definition is
\beq
\Pi^{\rho} (M^2)~\equiv ~\hat{L}\Pi^{\rho}(Q^2)~
\equiv~{\rm lim}_{n\rightarrow \infty}
{1 \over (n-1)!} (-1)^n(Q^2)^n({d\over {dQ^2}})^n
\Pi^{\rho} (Q^2) \label{borel},
\eeq
where the limit is understood in such a way that
\beq
n\rightarrow \infty,~~Q^2\rightarrow\infty,~~Q^2/n
\equiv M^2~{\rm is~fixed},
\eeq
and it is actually $M^2\sim m^2_{\rho}$ rather than $Q^2
\sim m^2_{\rho}$
that can be reached starting from large $M^2$.

In a somewhat simplified form the sum rules read
\beq
1+{{\alpha_s(M^2)}\over {\pi}}+
{{\pi^2}\over 3}{{\langle \alpha_s/\pi \cdot (G^2_{\mu\nu})^2\rangle}
   \over {M^4}}
+ {{448\pi^3}\over {81}}{{(\alpha_s^{1/2}\langle \bar q q\rangle )^2}
   \over {M^6}}+O(M^{-8})
\label{sr}
\eeq
$$~~~~~=
{{8\pi^2m^2_{\rho}}\over{g_{\rho}^2 M^2}}exp(-m^2_{\rho}/M^2)
+{2\over 3M^2}
\int_{s_0}^{\infty}exp(-s/M^2)R^{I=1}(s)ds,
$$
where $\alpha_s$ is the QCD coupling, $g_{\rho}$
the $\rho$ meson coupling ($= g_{\rho \pi \pi}$) related to the
 $e^+e^-$ width of the $\rho$ meson, $g_{\rho}^2/4\pi \approx 3$, and
   $\langle \bar q q\rangle$ and $\langle (G^2_{\mu\nu})^2\rangle$
   are the quark and the gluon condensates, respectively.
The integral on the right-hand side represents contribution of
the continuum.

The only observation concerning the sum rules which
is important for our purposes here is that even at $M^2\approx
m^2_{\rho}$ the
left-hand side of the sum rules (\ref{sr}) is calculable within the
short-distance approach to QCD, i.e., is dominated by the unit while
other terms
can be considered as small corrections. The sum rules agree with the
data, or the right-hand side, to within about 10 per cent.
Moreover, the $\rho$ contribution dominates over the continuum
at such $M^2$ to about the same accuracy. Eq.(\ref{sr})
can be considered, therefore, as a refined form of duality derived
within the fundamental QCD.

Now we come to the crucial point of the consistency of the effective
interaction
(\ref{one}) with the QCD sum rules. As is emphasized in the previous
section, it is far from being obvious that both the fundamental and
the effective Lagrangians
can be utilized simultaneously. In general, it is the opposite which
is true. It is just a very specific result that the QCD sum rules
apply even at
$M^2$ numerically close to $m^2_{\rho}$, which allows for a check of
self-consistency.
Indeed, if the cutoff $\Lambda_{UV}$ is as high as indicated by
eq.(\ref{numbers}), then there exists region of so to say moderate
$Q^2$,
\beq
0.5~{\rm GeV}^2~\le~ Q^2_{\rm moderate}~\le~ 1.5~{\rm GeV}^2,
\eeq
where the both approaches claim their validity.

Thus at least superficially we are allowed to add at $Q^2
\sim m^2_{\rho}$
the contribution of interaction (1) to the bare quark loop to get
\beq
\Pi^{\rho}(Q^2)_{\rm modified}~\approx~-{1\over {8\pi^2}}lnQ^2
\left(1~-~{{G_VQ^2}\over{2\pi^2}}lnQ^2\right)\label{new}
.\eeq
The first  term on the right-hand side here represents the bare quark
 loop which
is expected to dominate even at $Q^2\sim m^2_{\rho}$, as explained
 above. As for the piece proportional to $G_V$, its evaluation is
 in fact not unique, because
the effective interaction (\ref{one}) is non-renormalizable.
We could write, say, $G_V\Lambda_{UV}^2$ instead of
$G_VQ^2lnQ^2$. However, as far as we stretch eq.(\ref{new}) to
$Q^2\sim\Lambda_{UV}^2$, the two answers match each other smoothly.
Moreover,
we shall argue in the next section that in fact there is no much
uncertainty in our estimates, provided that the $\rho$ meson is
generated by the effective interaction (\ref{one}).

If we use the Borel transformed $\Pi^{\rho} (M^2)$, then the effect of
introduction of the new interaction is enhanced numerically:
\beq
\Pi^{\rho}(M^2)_{\rm modified}~\approx~{1\over {8\pi^2}}
\left(1~-~{{G_VM^2}\over{\pi^2}} \left(1-\gamma + ln M^2\right)
\right)\label{newM}
,\eeq
or
\beq
{{\delta\Pi^{\rho} (M^2)}\over {\Pi^{\rho}_0(M^2)}}~\approx~
{{G_VM^2}\over {\pi^2}}\left(1-\gamma+lnM^2\right),
\label{newm}
\eeq
where $\Pi^{\rho}_0(M^2)$ is the contribution of the bare loop graph
with massless quarks and $\gamma~\approx~0.577$ is the Euler constant.
Note that the effect of the new interaction grows with
$M^2$ as $M^2lnM^2$.
On the other hand, at large $M^2$ the effect should disappear because
of the onset of the asymptotic freedom. This emphasizes once more that
the effective interaction (\ref{one}) can be valid only at relatively
low momenta. Numerically, once we accept the estimates
(\ref{numbers}),
we can extrapolate (\ref{newm}) up to $M^2\approx\Lambda_{UV}^2$
and allow then for a form factor which would describe softening of
the effective interaction.

 For the new interaction to be consistent with the sum rules we
expect that (\ref{newm}) represents a small correction.
However, using eq.(\ref{new}), we conclude that introduction of
interaction (1)
results in fact in a considerable change of $\Pi^{\rho}(Q^2)$:
\beq
{{\delta\Pi^{\rho} (Q^2=m^2_{\rho})}
\over {\Pi^{\rho}_0(Q^2=m^2_{\rho})}}
{}~\approx~
-{{ G_Vm^2_{\rho}}\over {2\pi^2}}lnm^2_{\rho}~\sim~-0.25~lnm^2_{\rho},
\eeq
where $ln m^2_{\rho}~\approx~2$ if we take typically
$(250~ {\rm MeV})^2$ for the infrared
scale of the logarithm. This can hardly be compatible with the strong
dominance of the quark loop
established within the sum rules. The effect is even more dramatic,
if we turn to
the Borel transform $\Pi^{\rho}(M^2)$ pertinent to the sum rules (see
eq.(\ref{newm})):
\beq
{{\delta\Pi^{\rho} (M^2=m_{\rho}^2)}
\over {\Pi^{\rho}_0 (M^2=m_{\rho}^2})}~\sim~
0.5~(1-\gamma+lnm_{\rho}^2).
\eeq
We see that the new interaction cannot be considered as a small
correction at all but is to be rather iterated and summed up in the
spirit of the NJL model.

Thus, the conclusion is that the condition of consistency
with the QCD rules does not allow us to have $G_V$
as big as indicated by (\ref{numbers}). Because of the
importance of this conclusion we are going to discuss next
 to which extent it
depends on the particular parameterization (\ref{numbers}).

\section{Vector meson dominance vs sum rules}

In this section we will argue that the notion of $\rho$ meson
dominance at (Euclidean) momenta $Q^2\sim m^2_{\rho}$ is not indeed
 consistent with the QCD sum rules, so that the numerical
 contradiction found in the previous section is significant.

The point we wish to emphasize is in fact very simple. Namely,
consider $\Pi^{\rho}(Q^2)$ at $Q^2\sim m^2_{\rho}$. One can try to
apply either
$\rho$ meson dominance to evaluate $\Pi^{\rho}(Q^2)$ or just
 approximate it by the bare quark loop
graph. The former approximation assumes the distances to be large,
coupling large and the use of the language of strongly bound states.
The latter approximation rests on the assumption that the coupling is
already negligible and the language of free quarks is the appropriate
one. The both pictures
cannot be right when applied to the same quantity, $\Pi^{\rho}(Q^2)$
at the same values of $Q^2$. However, since the sum rules also imply
a kind of vector
meson dominance (VMD), we proceed to reiterate the argument on a
more technical level.

Let us start with the VMD picture, as it arises in the extended NJL
model.
In the limit of a large $N_c$ the correlator (\ref{corr}) is
approximated by the chain of the loop graphs (``bubble sum''):
\beq
\Pi^{\rho}(Q^2)
{}~\approx~{
{1\over {8\pi^2}}ln {\Lambda^2_{UV} \over Q^2+m^2_q}
\over {1+{{G_V Q^2}\over {2\pi^2}}ln {\Lambda^2_{UV}\over Q^2+m^2_q}}
},
\label{VMD}\eeq
where $m_q$ is the constituent quark mass, $m_q\approx 300~
{\rm MeV}$,
generated within the same NJL approach through the $G_S$ interaction
in (\ref{one}).

Neglecting $Q^2$ in the argument of the logarithm, one readily derives
the VMD for the correlator (\ref{corr}):
\beq
\Pi^{\rho}(Q^2)_{\rm VMD}~=~{1\over {g_{\rho}^2}}{{m^2_{\rho}}
\over{Q^2+m^2_{\rho}}}
\label{tw},
\eeq
with \beq
m^2_{\rho}~
=~{{2\pi^2}\over{G_Vln\Lambda^2_{UV}/m^2_q}}\label{ro}
\eeq
and
\beq
g^2_{\rho} = {8\pi^2 \over ln\Lambda^2_{UV}/m^2_q}.
\eeq
Upon applying the Borel transform (\ref{borel}) to (\ref{tw}), we
obtain
\beq
\hat{L}\Pi^{\rho} (Q^2)_{\rm VMD}
{}~=~{1\over {g_{\rho}^2}}{{m^2_{\rho}}\over {M^2}}
exp(~-m^2_{\rho}/M^2)\label{thir}.
\eeq
Eqs. (\ref{tw}) and (\ref{thir}) are expected to hold at low
$Q^2,M^2$, i.e.,
 much smaller than $\Lambda_{UV}^2$. Experimentally,
 eq.(\ref{thir}) holds at $M^2\le
m^2_{\rho}$ and this could be claimed a success of the
VMD model.

Now, the perturbative QCD proceeds in the following way. The
correlator $\Pi^{\rho}(Q^2)$ is approximated by the bare loop graph:
\beq
\Pi^{\rho} (Q^2)_{\rm pert ~QCD}~\approx~-{1\over {8\pi^2}}lnQ^2
\label{ft},
\eeq
where $Q^2$ is assumed to be ``large'', in contradistinction from
the VMD model.
Applying the Borel transform, one finds
\beq
\Pi^{\rho} (M^2)_{\rm pert~QCD}~\approx~{1\over {8\pi^2}}\label{fif}
\eeq
and this again holds experimentally to a reasonable accuracy at
$M^2\approx m^2_{\rho}$, since numerically (\ref{thir}) and
(\ref{fif}) are close to each other. This calculation could be
claimed a success of the short-distance approach to QCD.

Thus, the perturbative QCD derives (\ref{fif}) and treats (\ref{thir})
as a
phenomenological fit to the integral over $R^{I=1}(s)$, see
eq.(\ref{sr}).
In other words, one derives duality from the first principle,
provided that the existence of resonances is granted.
Within the effective
Lagrangian approach, on the other hand, one derives resonances.
The sum rules (\ref{sr})
become then a triviality, since both $\Pi^{\rho} (Q^2)$ in the
Euclidean region and $Im \Pi^{\rho}(s)\sim R(s)$ are dominated by
one and the same $\rho$ meson.

In terms of $\Lambda_{UV}$ the equation (\ref{tw}) is valid at
$\Lambda^2_{UV}\gg M^2$ (VMD case), while (\ref{ft})
is true if $\Lambda_{UV}^2\ll M^2$ (perturbative QCD case).
Moreover, if one evaluates the simplest quark graph as it is
prescribed by the VMD, then it
does not contribute to $\Pi^{\rho} (M^2)$ at all, since
\beq
\hat{L}({1\over {8\pi^2}}ln\Lambda^2_{UV})~=~0,
\eeq
while in case of perturbative QCD this graph dominates
(see eq.(\ref{fif})).

Thus, the success of the QCD sum rules implies necessity to modify the
VMD model around $M^2\sim m^2_{\rho}$ so that the composite nature of
the vector mesons would become manifest at such $M^2$.

To see whether the effect is dramatic numerically we should have
worked out an
interpolation between (\ref{fif}) and (\ref{thir}). A rigorous
treatment of this
transitive region seems to be out of reach of any known framework.
We can, however, suggest a simple-minded interpolation which might
reproduce gross features of the reality.

Namely, let us keep the $ln Q^2$ factor in eq.(\ref{VMD}).
Using the Borel transform (see eq. (\ref{borel})),
\beq
\hat{L}\{Q^{-2k}(lnQ^2/\Lambda^2)^{-\epsilon}\}~
\approx~{1\over {\Gamma(k)}}M^{-2k}
(lnM^2/\Lambda^2_{UV})^{-\epsilon},
\eeq
one obtains from (\ref{VMD})
\beq
\Pi_{\rho}(M^2)~\approx~{1\over {g_{\rho}^2}}{{m^2_{\rho}}\over{M^2}}
exp\left(-{{m^2_{\rho}}\over {M^2}}{{ln\Lambda^2_{UV}-lnm^2_q}
\over {ln\Lambda^2_{UV}-ln M^2}}\right),
\label{ei}
\eeq
where $\Lambda^2_{UV}>M^2$.

Eq.(\ref{ei}) demonstrates a remarkably sharp dissolution of $\rho$ as
$M^2$ approaches $\Lambda^2_{UV}$. For the purpose of eliminating the
disagreement with the sum rules one needs at $M^2\sim m^2_{\rho}$
\beq
{{ln \Lambda^2_{UV}-lnm^2_q}\over{ln \Lambda^2_{UV}-ln m^2_{\rho}}}
{}~\sim~3,
\label{nt}
\eeq
so that the effect of ``elementary'' $\rho$ meson generated via the
interaction (\ref{one}) goes away at Euclidean $Q^2\sim m_{\rho}^2$.
Although this conclusion  contradicts naive expectations based
on the VMD,
let us note that the VMD could still
be a valid approximation at Minkowski momenta and down to, say,
$Q^2\sim 0$.

To summarize,
success of the sum rules in the $\rho$ channel implies that the
VMD is to be replaced by the fundamental QCD
around $M^2\sim m^2_{\rho}$, which means in
turn a change in the fitting parameters (\ref{numbers}).
Our eq.(\ref{ei}) above is an attempt to change $\Lambda_{UV}$.
However, as we shall see in the next section, the low value
of $\Lambda_{UV}$ is not favoured by consideration
of the pseudoscalar, or the pion channel. Therefore it is worth
emphasizing that the most
straightforward solution to the problem
is to assume that $G_V$ is substantially smaller than that indicated
by (\ref{numbers}).
This would imply, however, that light vector mesons ($ m^2_{\rho} \ll
\Lambda^2_{UV}$) are not generated by
the effective
interaction (\ref{one}).

\section{Sum rules in pseudoscalar channel}

In this section we will argue that introduction of effective
interaction (\ref{one}) with values of $G_S$ and $\Lambda_{UV}$
as indicated by (\ref{numbers}) could resolve a long
standing problem of QCD-based phenomenology.

Namely, one can consider sum rules in the $\pi$ channel in exactly
the same way as
in the $\rho$ channel outlined above. The corresponding current is
defined as
\beq
j^{\pi}~=~{1\over 2}(\bar u i\gamma_5u-\bar d i\gamma_5d)
\eeq
and the sum rules take the form \cite{novikov}:
\beq
\left({{\alpha_s(M^2)}\over{\alpha_s(\mu^2)}}\right)^{8/9}
\left(1+\pi^2{{\langle \alpha_s/\pi\cdot
(G^a_{\mu\nu})^2\rangle }\over {3M^4}}+
{{896\pi^3}\over {81}}{{\langle \alpha_s^{1/2}\bar q q\rangle ^2}
\over {M^6}}+ O(M^{-8})\right)
\label{pi}
\eeq
$$={{16\pi}\over{3M^4}}\int ds~exp(~-s/M^2)Im \Pi^{\pi}(s),$$
where
$Im\Pi^{\pi}(s)$ is the imaginary part of the correlator of two
currents
$j^{\pi}$ and the factor $(\alpha_s(M^2)/\alpha_s(\mu^2))^{8/9}$ is
due to a non-vanishing anomalous dimension of $j^{\pi}$. As far as
$Im\Pi^{\pi}(s)$ is concerned, the only experimentally known
contribution to it comes from the pion:
\beq
Im\Pi^{\pi}_{\rm pole}~=~\pi f_{\pi}^2m_{\pi}^4(m_u+m_d)^{-2}
\delta(s-m^2_{\pi}),
\label{pole}\eeq
where $f_{\pi}\approx 93~ {\rm MeV}$ and $m_{u,d}$ are the current
 quark masses.

Now, it has been demonstrated that the sum rules (\ref{pi}) do not
hold
experimentally as they are stated \cite{novikov}. The point is that
the pole
contribution (\ref{pole}) alone, with negligence of the rest of
$Im \Pi^{\pi}
(s)$ which is positive definite, is large enough to break asymptotic
freedom in the pion channel at
\beq
(M^2)^{\pi}_{\rm crit}~\approx~2~{\rm GeV}^2
\label{crit}
.\eeq
Moreover, the actual number could be even higher, since we neglected
all other states with the same quantum numbers.
Note that the asymptotic freedom is represented by the unit in the
left-hand
side of eq.(\ref{pi}). The corresponding scale in the $\rho$ channel
discussed
in the previous section is numerically $(M^2)^{\rho}_{\rm crit}
\approx 0.6 ~{\rm GeV}^2$
and this was the basis for the whole success of the sum rules in
describing the $\rho$ meson.

In this way the sum rules reveal that the pion
is not dual to the quark loop, in contradistinction from the $\rho$
meson case. More specifically,
one can prove existence of a new contribution for $M^2$ in
the window $0.5~{\rm GeV}^2<M^2<~2~{\rm GeV}^2$.
At $M^2>2~{\rm GeV}^2$ the sum rules can
be dominated by the bare quark loop, while at $M^2<0.5~{\rm GeV}^2$
 the power like corrections to the sum rules become important
and the sum rules are no longer sensitive to new physics.
At $M^2=0.5~{\rm GeV}^2$ it is not less than the pion contribution.
while at $M^2 = 2~{\rm GeV}^2$ the new contribution is still larger
than 10\% of the pion contribution.

Thus the problem is that the power corrections accounted for in
(\ref{sr}) and
(\ref{pi}) fail to reproduce this change of scales with the switching
from the
$\rho$ channel to the $\pi$ channel \cite{novikov}. What we propose
here is to ascribe this difference to the presence of the
effective interaction (\ref{one}) in the $\pi$ channel.

At $M^2<\Lambda_{UV}^2$ the estimates of the effect of the new
interaction can be made similar to (\ref{new})~--~(\ref{newm}):
\beq
\Pi^{\pi}(Q^2)_{\rm modified}~\approx~{3\over {16\pi^2}}Q^2 lnQ^2
\left(1~+~{{3G_SQ^2}\over{4\pi^2}}lnQ^2\right)\label{newpi}
,\eeq
\beq
\Pi^{\pi}(M^2)_{\rm modified}~\approx~{3\over {16\pi^2}}M^2
\left(1~-~{{3G_SM^2}\over{\pi^2}}\left({3\over 2} -\gamma +lnM^2\right)
\right)\label{newpiM}
,\eeq
or
\beq
{{\delta \Pi^{\pi} (Q^2)}\over {\Pi^{\pi}_0 (Q^2)}}~\approx~
{{3G_SQ^2}\over {4\pi^2}}lnQ^2,
\eeq
\beq
{{\delta\Pi^{\pi}(M^2)}\over{\Pi^{\pi}_0(M^2)}}~\approx~
-{{3G_SM^2}\over{\pi^2}}\left({3\over 2} -\gamma+lnM^2\right),
\label{newmp}
\eeq
where by $\Pi^{\pi}_0 (Q^2)$ and $\Pi^{\pi}_0 (M^2)$ we understand
the contribution of the bare quark loop for the sake of normalization.

Eq.(\ref{newmp}) is supposed to be valid at $M^2$ much smaller than
$\Lambda_{UV}^2$. Applying it at $M^2=0.5~{\rm GeV}^2$, we find:
\beq
{{\delta\Pi^{\pi} (M^2=0.5~{\rm GeV}^2)}\over {\Pi^{\pi}_0(M^2=0.5~
{\rm GeV}^2)}}~\sim~2.5.
\label{large}
\eeq
Literally, the analysis of the sum rules suggests a new contribution
of order one in the same units (see above). The factor we get now is
in rough agreement with that estimate.

To get a better estimate we should have iterated the effect of the new
interaction, since its effect (\ref{large}) turned out to be large.
The change brought by the iterations of the effective interaction
is remarkably simple and well known within the NJL model. Namely,
within the NJL model the summation of the chain of
the graphs generated by the interaction (\ref{one}) produces a pion.
In this way we reproduce the contribution of the pion into the
right-hand side of the sum rules (\ref{pi}) and explain the failure of
the sum rules which do not account for interaction (\ref{one})
at $M^2\sim 0.5~{\rm GeV}^2$.

Thus the effective interaction in the pseudoscalar channel
with the parameters indicated in (\ref{numbers}) provides a natural
exlanation to the phenomenon found in Ref.\cite{novikov}.
Namely the four-quark interaction gives rise to the
pion as is proposed in the original papers \cite{nambu}. At $M^2\sim
\Lambda^2_{UV}\sim 1.5~{\rm GeV}^2$ the effective interaction is
dissolved and its contribution is replaced by the bare quark loop
which dominates in the region of small effective coupling of QCD.
The estimate of the mass scale of the onset of asymptotic freedom
in the pion channel as $2~{\rm GeV}^2$ (see above) turns out
to be in reasonable
agreement with the estimate of $\Lambda_{UV}^2$ within the NJL model.

Of course this picture does not explain by itself why the vector
and the pseudoscalar channels look different. Since the sum rules
are derived within the fundamental QCD, there should be new
corrections, not accounted for in the standard \cite{SVZ}
approach. This conclusion has been reached long time
ago \cite{novikov}. There were also speculations on
possible new sources of corrections \cite{novikov}. What is common
for these
corrections is that they depend on a high inverse power of $M^2$.
In conclusion of this section we would like to make a comment
that in fact the matching of the effective interaction (\ref{one})
with fundamental QCD looks most natural, if $M^{-2}$ corrections
are introduced. Indeed,
if we invoke $M^{-2}$ terms, then the difference
between the channels is least dramatic; namely, the
difference between $M^2_{\rm crit}$ in the $\pi$ and $\rho$
channels (2 ~${\rm GeV}^2$
and 0.5~ ${\rm GeV}^2$, respectively) is a factor of 4, which does not
look so
drastic by itself. What makes
the situation especially difficult for the standard sum rules is that
 they
operate with $M^{-4},M^{-6}$ corrections which boost the difference
to the factor
$\sim 10-10^2$ and, moreover, these corrections are well fixed
numerically and
do not allow for speculations on numbers. In Ref.\cite{novikov} an
attempt is
made to ascribe the difference to the contribution of direct
instantons. But
that contribution depends on $M^2$ even more drastically,
like $M^{-9}$
and the difference between the two channels is difficult to explain.

The possibility of existence of $M^{-2}$ corrections has been realized
only
recently \cite{VZrecent,Brown}. In the most direct way they arise
from the
so-called ultraviolet renormalons \cite{thooft}. Ultraviolet
renormalons are
associated with insertions of vacuum bubble insertions into a gluon
line.
An extra bonus of such interpretation is that the gluon line carries
a large virtual momenta $k^2$,
\beq
k^2~\sim~e^nQ^2,
\eeq
where $n$ is the order of perturbative calculation considered and $e$
is the base of natural logs. In general, gluon exchanges generate
various
effective interactions but it was demonstrated that four-fermion
operators (\ref{one}) dominate at large $n$ and induce $M^{-2}$ terms
\cite{vainshtein}.
Once we take $M^2$ to be low enough, the language of asymptotic
freedom is no longer
valid and the four-fermion operators governing the renormalon
become effective interaction (\ref{one}).
\section{Conclusions}

To summarize, confronting the NJL model with QCD sum rules uncovers
a nontrivial
interplay of the two approaches. In the $\rho$ channel the success
of the
QCD sum rules \cite{SVZ} calls for a revision of the NJL model for
vector
mesons. In the $\pi$ channel, to the contrary, the QCD sum rules badly
 need a new
contribution \cite{novikov} and we have demonstrated that the NJL
effective
interaction could well be used for this purpose.

The only way how this effective interaction could be accommodated into
the perturbative
 QCD seems to be through ultraviolet renormalon \cite{vainshtein}.
(Such an interaction might also be induced by the nonperturbative
effects of QCD \cite{kondo}.)
Such a
hypothesis assumes, however, that the renormalon contribution is large
numerically. Let us note that similar assumption in fact underlies the
 original
QCD sum rules as well. Indeed, one can argue that $M^{-4}$ corrections
arise from divergencies of perturbation theory in large orders and are
related to the so-called infrared renormalon \cite{mueller}. However,
it is not obvious at all that these corrections are numerically
important and are not screened by lower order perturbative terms.
The basis of the phenomenology of the sum rules is the assumption
that the $M^{-4}$ corrections are numerically large.

 For the ultraviolet renormalon to be relevant, there should be a large
numerical factor as well. However, what makes the phenomenology of
$M^{-2}$
corrections still much less definite is that there is no way
to relate the $M^{-2}$ terms in different channels, even if one is
prepared
to assume that these terms are enhanced numerically.
It is at this point that the phenomenology of the NJL model could
play a crucial role.

Indeed the chain of the arguments above gets closed through the
prediction
\beq
G_V~\ll ~G_S,
\label{pred}
\eeq
where $G_{V,S}$ are the constants of the effective four-four-fermion
interactions
(\ref{one}). We need (\ref{pred}) to ameliorate the sum rules
in the pseudoscalar channel without destroying them
in the vector channel.

Although eq.(\ref{pred}) contradicts the spirit of the extended NJL
model (see, e.g., review in Ref.\cite{volkov}),
it is not obvious that
the prediction (\ref{pred}) can be ruled out phenomenologically.
In particular, the most
elaborated comparison of the NJL model with the experimental data
performed
in Ref.\cite{bijens} reveals that the solution with $G_V=0$
gives a very satiosfactory fit to all known parameters of the
low-energy pion interactions. Moreover, any $G_V\neq 0$ drives the
predicted value of the constant $g_A$ governing the beta-decay of
neutron
off its experimental value. Less dramatically, taking $G_{V}=0$
improves
fits to some other parameters as well (see \cite{bijens}).
 Furthermore, as is noted in \cite{bijens}, the NJL model with $G_V=0$
is equivalent to the effective QCD Lagrangian of Ref.\cite{espriu}
which turned out to be successful in other phenomenological
applications
\cite{pich}.

Thus it seems fair to say that the NJL model with $G_V=0$ and
$G_S\neq0$
results in a sound phenomenology, although it puts pseudoscalar and
vector
meson on different footing, as is required by the QCD sum rules.

{\bf Acknowledgements}

The authors are thankful to G. Grunberg and M.K. Volkov for useful
 discussions. One of the
authors (V.I.Z.) acknowledges warm hospitality extended to him during
his stay
with the theory group at Nagoya University when this work was started.
Part of this work was done while K.Y. was at Institute for Theoretical
 Physics, University of California at Santa Barbara (Workshop
 ``Weak Interactions'').
This work was supported in part by the U.S. Department of Energy, and
 by a Grant-in-Aid for Scientific Research from the Japanese
Ministry of Education, Science and Culture (No. 05640339) and the
Ishida Foundation.

\end{document}